# High Throughput Separation of Cells Achieved Through the Particle Characteristic Dielectrophoretic Response and Further Focusing

Ngoc-Duy Dinh[1*], Dinh-Tuan Phan[2]

[1]Center for Misfolding Diseases, Department of Chemistry, University of Cambridge, The Old Schools, Trinity Ln, Cambridge CB2 1TN
Email: dd552@cam.ac.uk; dinhngocduy@u.nus.edu

[2]Institute for NanoBioTechnology, Johns Hopkins University, 3400 North Charles, Croft Hall 100 Baltimore, MD 21218

**ABSTRACT**

Previous devices to separate cells by the characteristic force they experience due to dielectrophoresis, which depends on the size and electric properties of the particle, were limited by the flow rates and particle concentrations separation could be achieved at. To unlock the potential of Lab-on-a-chip technology to create flexible, efficient and multifunction devices at low cost it is necessary to increase the rates at which separations can be performed. Here we present a device capable of high throughput continuous separation of cells from microparticles and demonstrate separation at 7μL/min in a microchannel with a high density of cells and microparticles in the sample of $10^8$-$10^9$ cells/mL. This device uses the characteristic response of the particles due to dielectrophoresis to provide the initial separation before crucially focusing the particles into completely separate particle streams allowing the operation at high flow rates and particle concentrations. The device is demonstrated with the separation of polystyrene (PS) beads 8 μm, 25μm and human dermal microvascular endothelial cell line (HMEC-1 cells) and human liver cell line (HepG2 cells) suspended in a DEP buffer solution. We show that it is possible to achieve not only high flow rate separation but also high sample purity with, after separation and focusing, one output containing 100% cells.

**KEYWORDS:** Cell separation, cell focus, dielectrophoresis, lab-on-Chip, manipulation

# Introduction:

The separation of cells and microparticles has many applications in biological and medical analysis where a fast sample preparation process is necessary. Conventional Fluorescence-Activated Cell Sorters (FACs machines) are capable of very high speed separation of bioparticles however there is increasing interest in developing a technique that can be miniaturised into a Lab-on-a-Chip device. Several techniques have been investigated including optical [1], mechanical (hydrodynamic [2-4] and acoustic [5]), magnetic [6]), electrical (dielectrophoresis [7-9], AC electroosmosis [10] and electrothermal [11]) and optoelectronic [12]. Dielectrophoresis (DEP) is commonly used for the manipulation and separation of dielectric particles. Particles experience either an attractive or repulsive force when placed in a non-uniform electric field created by nearby electrodes. The separation of cells and particles has been reported based on the size of particles [13-17] or other characteristics of particles that produce a different dielectrophoresis force [18-20]. In this report, we demonstrate a DEP device with high efficiency separation and focusing of cells and particles. The high throughput chip separates and focuses cells and particles based on the different characteristics of particles, with a high flow rate in a microchannel and with a high density of particles in the solution.

# Materials and Methods:

Device design: Our device design is illustrated in Figures 1(a) and 1(b). The DEP chip has two electrode systems: an electrode system for separating and another one for focusing the particles after separation. There are three inlets on the microfluidics chip, inlet #1 and inlet #3 are for the DEP buffer solution and for creating a hydrodynamic focusing of the cells and particles before separation. Cells and particles will go to the right side of the microchannel staying a distance from the channel wall to decrease the effects of the wall of the channel [8]. The inlet #2 is for cells and particles in DEP buffer suspension. There are three outlets, the first one for beads, the second one is waste and the last one is for cells. We use tubes to collect the samples allowing us to calculate the separation percentage. Figure 1(c) shows the operation principle of the device. The cells in the microfluidic channel are deflected by the combination of hydrodynamic force and DEP force. The first force is a hydrodynamic force from the syringe pump which drives the particles through the microchannel. The second force is a DEP force due to the electrodes and this force is perpendicular to the electrodes. The combination of the two forces deflects cells from the left to the right of the microchannel. Here we describe the operation, design, simulation, fabrication and bio-experiment of our chip and report on its separation and focus performance.

Numerical Simulation: The electric field distribution and its gradient were modelled numerically in CFD-ACE+ (CFD Research Corp, Huntsville, AL). We divide the simulations of electric field into two parts: one is the electrode system for separation and another is electrode system for focusing. First of all, the model of electrodes was established and then an AC electric field of 6 Vp-p was applied to the separation electrode system. Figure 2(a) shows the root mean square of ac electric field on XY-plane (Z = 0µm). Obviously there is a region of high electric field intensity around the separation electrodes. Figure 2(b) shows the distribution root mean square of the electric field above the separation electrode system, the field is plotted against x at three different values of y; 250 µm, 375 µm and 75 µm. The simulation root mean square of electric field above the focusing electrodes is shown in Figure 2(c) with 10 Vp-p. This figure shows where the maximum gradient electric field is, where the bioparticles experience the highest DEP forces. Figure 2(d) shows the distribution root mean square of electric field of above the focus electrode system versus X-axis at three different positions Y= 0 µm, Y= 175 µm, Y= 250 µm. The device fabrication process is shown in Figure 3: To fabricate the electrode system we used a pyrex wafer and deposited 3000 Å of titanium (Ti) by e-beam evaporation (step 1) which was then patterned by a lithography and etching process (step 2). The electrode width and spacing was 50 µm, 50 µm respectively and with a $45^0$ angle to the channel walls. The microchannel was fabricated by soft lithography techniques (step3) [21]. A PDMS replica was molded on an SU-8 and silicon master (SU-8 2015, Microchem, Newton MA) to create a microchannel with a width of 500 µm and a height of 40 µm (step 4). The PDMS replica was drilled with holes for reagent injection. Finally, we bonded the microchannel with a substrate by enhanced oxygen plasma treatment (step5).

Cell culture and sample preparation for bio-experiment: The sample was a mixture including HMEC-1 cells (HepG2 cell) and polystyrene beads 8 μm and 25μm in diameter suspended in a DEP buffer solution (55% sucrose [wt/vol], 0.3% glucose [wt/vol] and 0.735% RPMI [wt/vol]), 70 μS/cm conductivity with a final high concentration $10^8$ to $10^9$ cells/mL.

Cell culture: HMEC-1 cells, HepG2 (ATCC, HB 8065) were maintained at $37^0$C with 95%/5% air/CO2 in Fscove's modified Dulbecco's medium (DMEM, Gibco-BRL, NY) containing 10% (v/v) heat inactived feta bovine serum (FBS, Biological Industries, Israel) and an antibiotic which was prepared and maintained in M200 medium supplemented with low serum growth supplement (LSGS, both from Cascade Biologics Inc). The cells were then separated and put in a tube and the medium was changed to DEP buffer.

Experiment setup: We used three Hamilton Gastight syringes (Hamilton; Reno, NV) for three inlets 1mL: 1mL: 3mL, connected to a syringe pump (WPI SP2301W) to provide the sample and DEP buffer for the chip. A function generator (B3120A, 15 MHz Function/ Arbitrary waveform Generator, Agilent Technologies) was used to generate sinusoidal wave excitation up to 10 MHz, 10Vp-p. The target particles and cells were examined using a microscope (Olympus BX51TRf) on an optical table. The images and movies were recorded using a digital CCD camera controlled by a computer.

## Theory:

Dielctrophoresis was coined by H.A.Pohl [22] who performed an important experiment with small plastic particle suspension in insulating dielectric liquids [23, 24]. The particle would move in response to the application of a non-uniform electric field. Dielectrophoresis is combination of phoresis Greek mean force and dielectric [25]. It describes the force exerted on uncharged dielectric particles by virtue of their polarizability [26]. DEP force effect on particle in microchannel can express by x direction and y direction:

$$F_{xDEP} = \pi\varepsilon_m r^3 \text{Re}[f_{CM}]\frac{\partial}{\partial x}|E|^2 \tag{1}$$

$$F_{yDEP} = \pi\varepsilon_m r^3 \text{Re}[f_{CM}]\frac{\partial}{\partial y}|E|^2 \tag{2}$$

Where: $\varepsilon_m$ is the permittivity of the suspending medium, r is the radius of cell or particle, $\frac{\partial}{\partial x}|E|^2$ and $\frac{\partial}{\partial y}|E|^2$ are the electric filed gradient of x direction and y direction, $f_{CM}$ is the Clausius –Mossotti factor given by:

$$f_{CM} = \frac{\overline{\varepsilon}_p - \overline{\varepsilon}_m}{\overline{\varepsilon}_p + 2\overline{\varepsilon}_m} \tag{3}$$

Where $\overline{\varepsilon}_p$, $\overline{\varepsilon}_m$ are the complex permittivity of particles and medium respectively. The complex permittivity is defined as follows:

$$\overline{\varepsilon} = \varepsilon - j\frac{\delta}{\omega} \tag{4}$$

$\varepsilon$ is the permittivity, $\delta$ is the conductivity, $j^2 = -1$, $\omega$ is the angular frequency, $\text{Re}[f_{CM}]$ the real part of Clausius-Mossotti factor is described as follows:

$$\text{Re}[f_{CM}] = \frac{(\delta_p - \delta_m)}{(1+\omega^2\tau^2)(\delta_p + 2\delta_m)} + \frac{\omega^2\tau^2(\varepsilon_p - \varepsilon_m)}{(1+\omega^2\tau^2)(\varepsilon_p + 2\varepsilon_m)} \tag{5}$$

$\tau$ is the Maxwell-Wagner constant. For cells, a single shell model:

$$\overline{\varepsilon}_p = \overline{\varepsilon}_{mem} \cdot \frac{\gamma^3 + 2\left(\frac{\overline{\varepsilon}_i - \overline{\varepsilon}_{mem}}{\overline{\varepsilon}_i + 2\overline{\varepsilon}_{mem}}\right)}{\gamma^3 - 2\left(\frac{\overline{\varepsilon}_i - \overline{\varepsilon}_{mem}}{\overline{\varepsilon}_i + 2\overline{\varepsilon}_{mem}}\right)} \tag{6}$$

$\overline{\varepsilon}_p$, $\overline{\varepsilon}_{mem}$ are complex permittivity of the cytoplasm and membrane, d is cell membrane thickness, $\gamma = \dfrac{r}{r-d}$, r is radius of particles.

Cells and particles moving in microchannel are applied by a hydrodynamic force in X direction and Y direction:

$$F_x drag = 6\pi\eta r W_x \dfrac{dx}{dt} \tag{7}$$

$$F_y drag = 6\pi\eta r W_y \dfrac{dy}{dt} \tag{8}$$

Where: $\eta$ is the viscosity of medium, $W_x$ and $W_y$ are the wall correction factor in the x and y direction respectively. dx/dt, dy/dt are the instantaneous cell velocities in the x and y direction, respectively. The total force acting cells or particles moving in microchannel can be described as:

$$\vec{F} = \vec{F}_{DEP} + \vec{F}_{drag} + \vec{F}_{grav} + \vec{F}_{bouy} \tag{9}$$

Where: $\vec{F}_{grav}$ is gravitational force, $\vec{F}_{bouy}$ is buoyancy force. At steady state, the DEP force is balanced by the viscous drag:

$$\vec{F}_{DEP} + \vec{F}_{drag} = \vec{0} \tag{10}$$

Thus, the velocity of cell or particle follow x direction and y direction as:

$$v_x = \dfrac{\varepsilon_m r^2}{6\eta W_x} \text{Re}[f_{CM}] \dfrac{\partial}{\partial x}|E|^2 \tag{11}$$

$$v_y = \dfrac{\varepsilon_m r^2}{6\eta W_y} \text{Re}[f_{CM}] \dfrac{\partial}{\partial y}|E|^2 \tag{12}$$

## Results and Discussions:

**Experiment:** The separation and focusing of cells and particles is based on the physical characteristics of the particles. We set up the experimental parameters as: frequency 100 kHz, 10 V peak- peak and flow rate 2 μL /min. At these values the cells are deflected from left channel to right channel. Cells are separated from the mixed sample of beads 8 µm, beads 25 µm and HMEC-1 cells. At these conditions the cells are deflected rapidly as shown in Figure 4(a). The white dots are cells and the black dots are beads. After the cells and beads are separated, they are focused by the focusing electrode system as shown in Figure 4(b). The Figure 4(c) is an inverted fluorescent image of cells deflection in a high flow rate of 7µL/min. The focusing of cells and particles is very important for optical detection [27, 28] or flow cytometry [29]. However, in this report, we used our focusing of particles for collecting separated species of microparticle. The cells and particles are collected and the percentage of cells and beads are calculated at the three outlets (Figure 4d).

High throughput separation and focusing: Lab-on-chip devices focus on decreasing the cost and increasing the efficiency of devices. The advantage of our chip is high throughput separation and focusing. We have designed an electrode separation array which is 1.5 mm in length. When we operate with flow rate 2µL/min, cells are deflected rapidly at the beginning of the electrode. When we increase the flow rate to 4 µL/min, 5 µL/min and 7 µL/min (Figure 5a), cells are deflected further onto the electrode. The electrode array is designed to make sure all of cells are deflected after passing over the separation electrode system. Previously reported DEP devices for the separation of cells achieve only limited flow rate [18]. Here we demonstrated a DEP chip for the separation and focusing of cells at flow rates over 5 µL/min. Figure 5(b) illustrates the relationship between the cell separation efficiency of the device and the flow rate at frequency 100 kHz and 10 V peak- peak. We use a sample with a mixture of cells in a high density suspension of $10^8$-$10^9$ cells/mL [18, 20]. Figure 5(c) shows how the high density mixture is separated rapidly when passed over the separation electrode array. We can see all of the cells deflect to the right of microchannel. The separation can be controlled by the changing the ratio of the hydrodynamic force and the DEP force, hydrodynamic force is generated by the syringe pump (flow rate). DEP force is controlled by function generator (voltage and frequency). We reach the maximum separation percentage by controlling the frequency, voltage and flow rate. However, high voltage could damage cells. In our experiment we use a limit to voltage of 10 V peak-peak which improves cell viability [30, 31]. Figure 5(d) illustrates the relationship of separation efficiency with voltage and frequency. The voltage applied has a strong effect to the DEP force which can be used to control the separation efficiency at a desire frequency. We can see a maximum separation efficiency occurred at 10 V peak-peak and 100 kHz frequency. The angle of the electrodes also affects their ability to deflect cells [32]. In our chip, the electrode is designed with an angle $45^0$ as shown in Figure 1(c).

**Discussion:**

The miniaturizing lab on chip [33-35] devices have been shown the advantages in biomedical applications such as for neuron culture or generated microdroplet [36-40]. Lab on Chip has demonstrated to revolute FACs machines in separation. DEP separation is based on the electrical properties of cells and particles. It is necessary to improve and control many parameters to obtain efficient separation. One parameter which is very important for a commercial product is high throughput. In this paper we show how this can be achieved with contactless DEP (cDEP) to separate cells and particles based on the characteristics of the particles, see supplementary videos. The width of the electrodes and the gap between electrodes is enough to generate the electrical fields necessary for to deflection. If the gap is too small, it will lead to excessive joule heating and electrothermal convection [18], if too large the fields will be too weak to deflect the particles. This paper shows the successful separation and focusing of cells from beads, next it will be necessary to assess the viability of the cells after separation. The cells may be damaged by the electric fields or simply from being suspended in the DEP buffer solution for too long. Thus, integration of a device capable of exchanging the medium the cells are suspended in along with a separation device may prove beneficial.

**Conclusion:** We have presented a novel high throughput device for the separation and focusing of cells and microparticles by their characteristic dielectrophoretic forces. The device was demonstrated at a high flow rate 7μL/min in microchannel with a high density of cells and microparticles in the sample $10^8$-$10^9$ cells/ml. We have demonstrated the successful separation of cells from microparticles achieving a high purity of cells 100 %. This show the potential of this technique for the isolation of samples for the detection of different species of microparticles and cells in a low cost, high efficiency, automated diagnostic device.

**REFERENCES**

1. MacDonald, M. P; Spalding, G. C; Dholakia, K. Nature. 2003, 426, 421– 424
2. Nicole, P. *Lab Chip*. **2007**, 7, 1644 – 1659.
3. Hsu, C.H.; Di Carlo, D, Chen C.; Iramia, D.; Toner, M. *Lab Chip*. **2008**, 8, 2128 – 2134.
4. Mao, X.; Lin, S. S. C.; Dong, C.; Huang, T. J. *Lab Chip*. **2009**, 9, 1583 – 1589.
5. Shi, J.; Ahmed, D.; Mao, X.Lin, S. C.; Lawit, A.; Huang, T. J. *Lab Chip*. **2009**, 9, 2890 – 2895.
6. Adams, J. D.; Kim, U.; Soh, H. T. *PNAS*. **2008**, 47, 18165–18170.
7. Pohl, H. A.; Hawk, I. *Science*. **1966**, 3722, 647 – 649.
8. Jason, G. K.; Michael, T. W. L.; Martin, A. S.; Klavs, F. J. *Anal. Chem*. **2006**, 78, 5019–5025.
9. Braschler, T.; Demierre, N.; Nascimento, E.; Silva, T, Oliva, A. G, Renaud, P. *Lab Chip*. **2008**, 8, 280-286.
10. Kuo, C. T.; Liu, C. H. *Lab Chip*. **2008**, 8, 725 – 733.
11. Lian, M.; Islam, N.; Wu, J. IET Nanobiotechnol. **2007**, 1, 3.
12. Chiou, P. Y.; Ohta, A. T.; Wu, M. C. *Nature*. **2005**,436, 370-372
13. Chang, S.; Cho, Y. H. *Lab Chip*. **2008**, 8, 1930-1936.
14. Han, K. H.; Frazier, A. B. *Lab Chip*. **2008**, 8, 1079-1086.
15. Han, K. H.; Han, S. I.; Frazier, A. B. *Lab Chip*. **2009**, 9, 2958-2964.
16. Cui, H. H.; Voldman, J.; He, X. F.; Lim, K. M. *Lab Chip*. **2009**, 9, 2306-2312.
17. Zhu, J.; Tzeng, T. R.; Xuan, X. *Electrophoresis*. **2010**, 31, 1382-1388
18. Vahey, M. D.; Voldman, J. *Anal. Chem*. **2009**, 81, 2446–2455.
19. Hakoda, M.; Wakizaka, Y.; Hirota, Y. *Biotechnology Progress*, 2010
20. Shafiee, H.; Sano, M. B.; Henslee, E. A.; Caldwell, J. L.; Davalos, R. V. *Lab Chip*, **2010**, 10**,** 438-445
21. Xia, Y.; Whitesides, G. M. *Angew. Chem. Int. Ed. Engl.* **1998**, 37, 551-575
22. Pohl HA, ''Dielectrophoresis'', New York, Cambridge University Press; 1978

23. Morgan H, Hughes MP and Green NG, ''Separation of submicron bioparticles by dielectrophoresis'', Biophys J .1999, 42, 279-293

24. Morgan, H., Green, N. and Sun, T, ''Electrokinetics of Particles and Fluids.'' In: *Microtechnology and Nanotechnology in Biomedical Applications*, Oxford University Press. (In Press) 2008

25. Jones TB, ''Electromechanics of Particles'', New York: Cambridge University Press; 1995

26. Morgan, H., Green, N. and Sun, T, ''Electrokinetics of Particles and Fluids.'' In: *Microtechnology and Nanotechnology in Biomedical Applications,* Oxford University Press. (In Press) 2008

27. Holmes, D.; Morgan, H.; Green, N. G. *Biosensors and Bioelectronics.* **2006**, 21, 8, 1621-1630.

28. Lee, G. B.; Lin, Y. H.; Lin, W. Y.; Guo, T.F. *Transducers* 2009, Denver, CO, USA, **2009**, 2135-2138.

29. Lin, H.; Lee, G. B.; Fu, L. M.; Hwey, B. H. *J. Microelectromech. Syst.* **2004**, 13, 923-932.

30. Lin, R. Z.; Ho, C. T.; Liu, C. H.; Chang, H. Y. *Biotechnology Journal.* **2006**, 1, 949-957.

31. Ho, C. T.; Lin, R. Z.; Chang, W. Y.; Liu, C. H. *Lab Chip*. **2006,** 6**,** 724 – 734.

32. Yunus, N. A. M.; Green, N. G. *Electrostatics. 2007*, 142, 12068.

33. Dinh, N. D., Chiang, Y. Y., Hardelauf, H., Baumann, J., Jackson, E., Waide, S., ... & Peyrin, J. M. *Lab on a Chip*. **2013**, 13(7), 1402-1412.

34. Dinh, N. D., Chiang, Y. Y., Hardelauf, H., Waide, S., Janasek, D., & West, J. JoVE (Journal of Visualized Experiments). **2014** May 20(87):e51389.

35. Phan DT, Jin L, Wustoni S, Chen CH. Lab on a Chip. 2018;18(4):574-84.

36. Dinh, N. D., Kukumberg, M., Nguyen, A. T., Keramati, H., Guo, S., Phan, D. T., ... & Kofidis, T. Lab on a Chip. **2020**, 20(15), 2756-2764.

37. Dinh, N. D., Luo, R., Christine, M. T. A., Lin, W. N., Shih, W. C., Goh, J. C. H., & Chen, C. H. *Small*. **2017** 13(24), 1700684.

*38.* Phan DT, Nguyen NT. *Applied Physics Letters.* **2014** Feb 24;104(8):084104.

39. Luo, R., Dinh, N. D., & Chen, C. H. *Biomicrofluidics*. **2017**, 11(3), 034107.

40. Luo, R., Wu, J., Dinh, N. D., & Chen, C. H. (Adv. Funct. Mater. 47/2015). *Advanced Functional Materials*. **2015**, 25(47), 7245-7245.

# ACKNOWLEDGMENT

We also thank members of Bioelectronics group and Micro-system and Control Laboratory for helpful comments.

**Figures Caption**

Figure 1. (a) Design of the device: The chip is designed with two electrode systems: an electrode system for separating and an electrode system for the focusing of particles after separating. The chip is designed with three inlets. Inlet #1 and inlet #3 are for DEP buffers and create hydrodynamic focusing of particles before separation. Inlet #2 is for bio-particles. The device has three outlets. The first one is for beads the second one is waste, and final one is the cells outlet. (b) Shows 3-Dimension figure of device. (c) Shows the operation principle of the device.

Figure 2. (a) Shows the square of the electric field of the separation electrodes in the XY-plane (Z = 0μm). (b) Shows square of electric field of the separation electrode versus X-axis (three positions Y= 250 μm, Y= 350 μm, Y= 75 μm). The simulation of the electric field of the focusing electrode system is shown as (c) with 10 Vp-p. (d) Shows the distribution of the electric field of the focusing electrode system versus X-axis at three different positions Y= 0 μm, Y= 175 μm, Y= 250 μm.

Figure 3. Schematic diagram of the Chip fabrication process: The micro-fabrication process used IC fabrication techniques to fabricate the electrode: first, Ti coating on glass wafer, lithography and wet etching, to create the electrode. Next we used soft lithography to fabricate the microchannel. We coated SU8 on a silicon wafer and defined a master with photolithography. Then by pouring PDMS, drilling inlet holes and bonding with oxygen plasma treatment the channel was created.

Figure 4. (a) Illustrates cells being separated on our chip: The experiment is demonstrated with 2 µL/min flow rate, 100 kHz frequency, 10 Vp-p. The white dots are cells and black dots are beads. We can see cells deflected from the left side of the microchannel to the right side of the microchannel. The scale bar is 100 µm. (b) Illustrates cells being focused on our chip. Cells and particles are focused into a straight line by the focusing electrode system. The scale bar is 100 µm. (c) Illustrates the inverted fluorescence image of cell deflection in a high flow rate of 7µL/min. The scale bar is 100 µm. (d) Shows the separation percentage of cells and beads at the three outlets with flow rate 2µL/min, 100 kHz frequency, 10 Vp-p: the purity of the separated suspension of cells is assessed at inlet #3.

Figure 5. (a) Illustrates cells deflected on the chip with a high flow rate of 7µL/min. The scale bar is 100 µm. (b) Cell Separation efficiency of device at 100 kHz and 10 Vp-p for flow rates of 2, 4, 5,7,10 and 15 mL/min, When we increase the flow rate to 15 mL/min the efficiency of separation drops to just 10%. The scale bar is 100 µm. (c) Illustrates the ability of the device to deflect cells in high density suspensions. Factor that effect the device's ability to separate and focus particles include voltage and frequency. (d) Illustrates the separation efficiency at flow rate 2 µl/min of the device at 1, 10, 50 and 100 kHz as voltages increase from 3 Vp-p to 10 Vp-p.

# Figures:

**Figure 1**

(a)

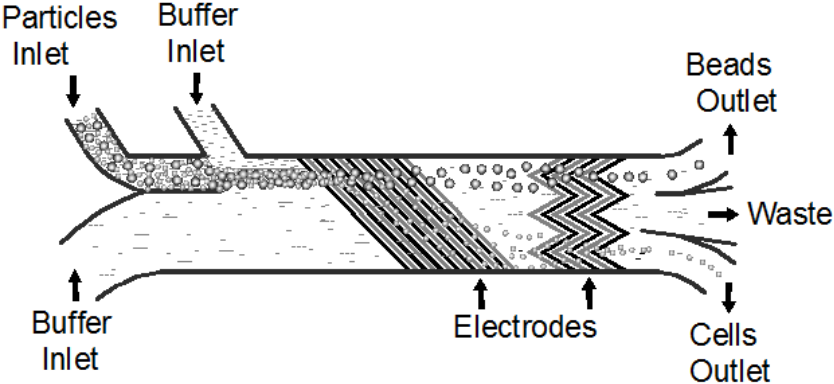

(b)

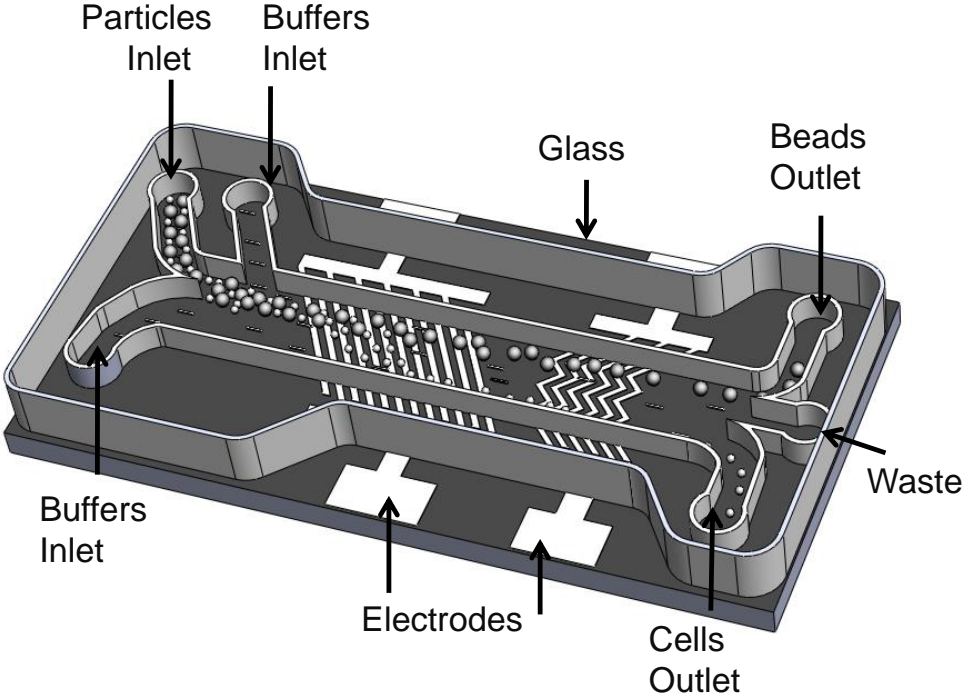

**(c)**

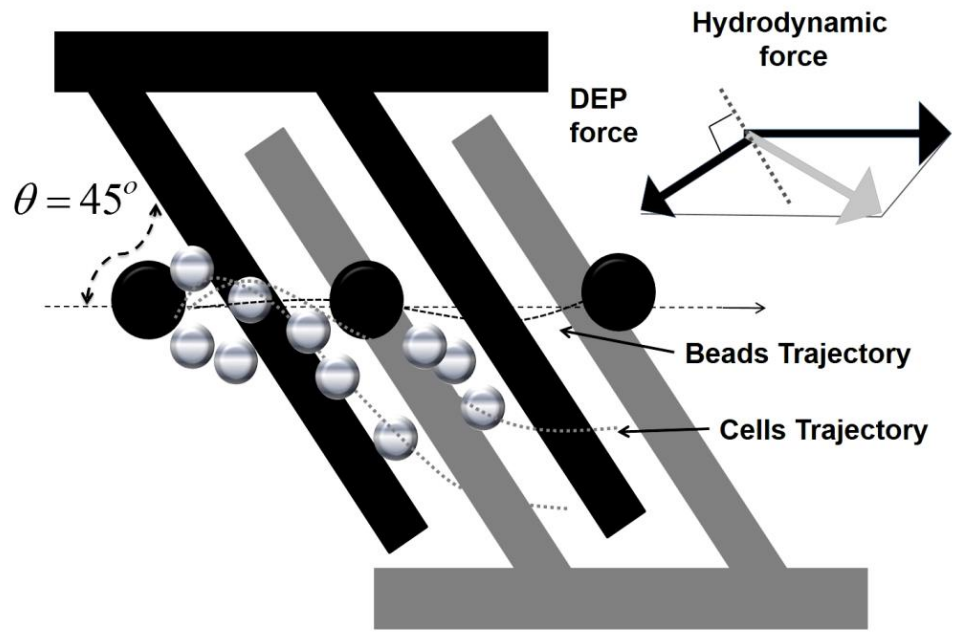

**Figure 2**

**(a)**

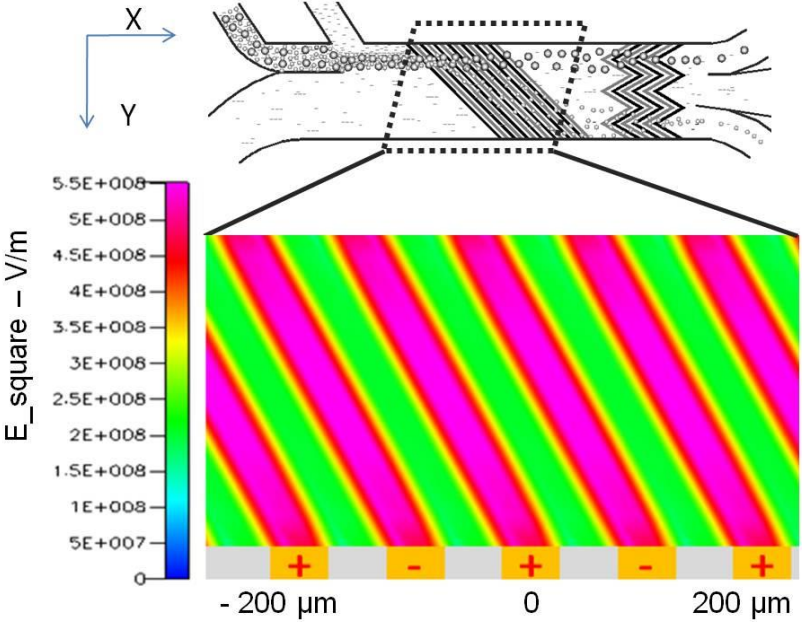

**(b)**

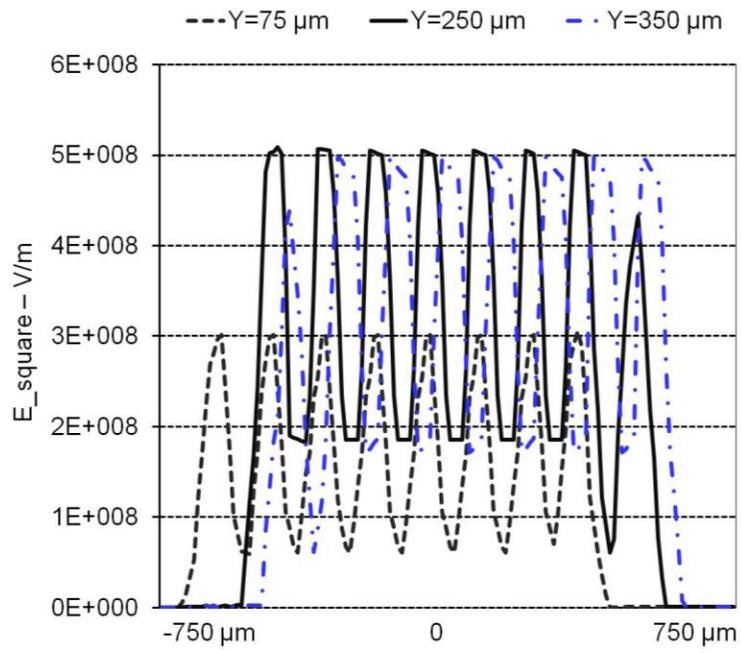

**(c)**

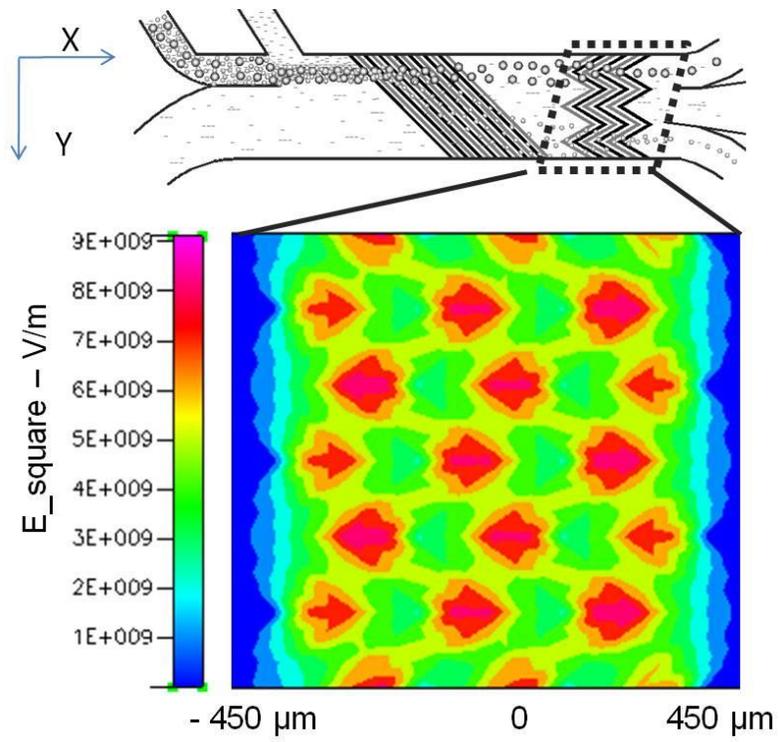

**(d)**

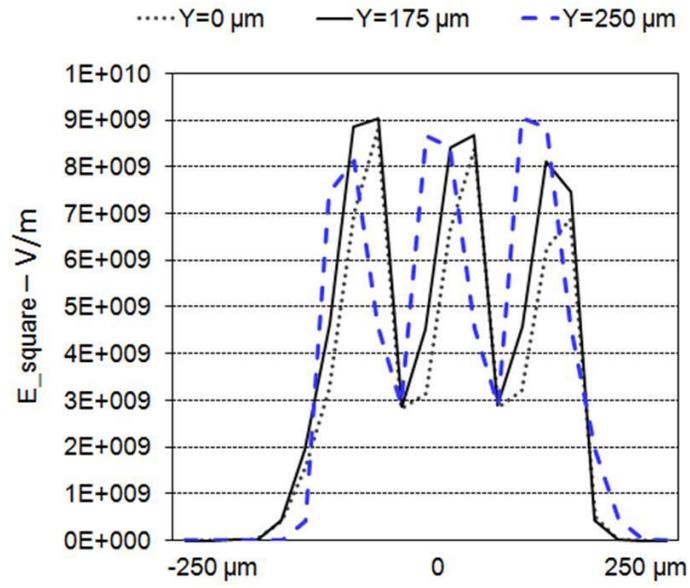

**Figure 3**

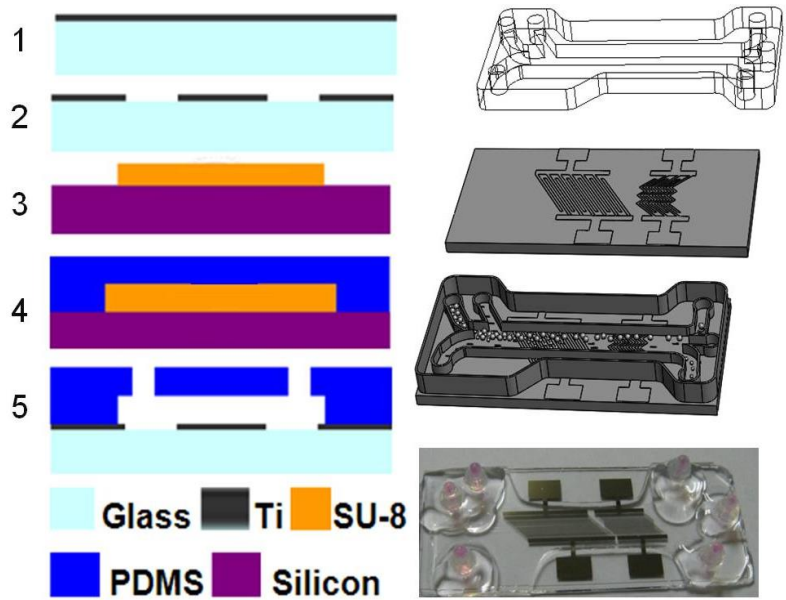

**Figure 4**

(a)

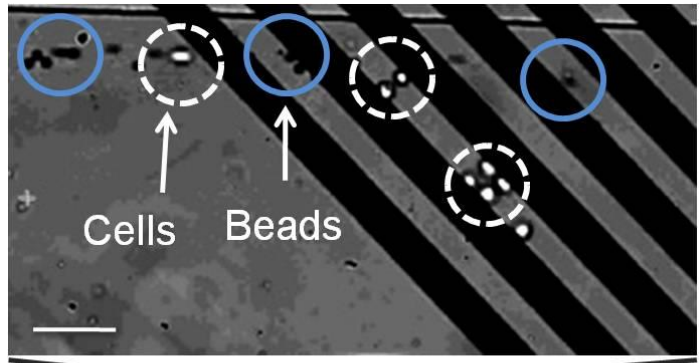

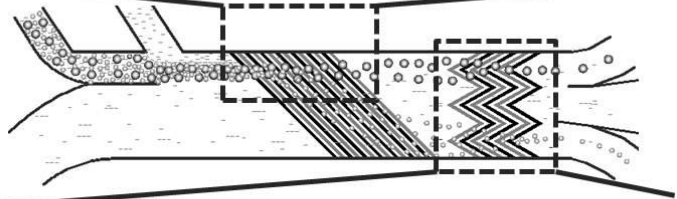

**(b)**

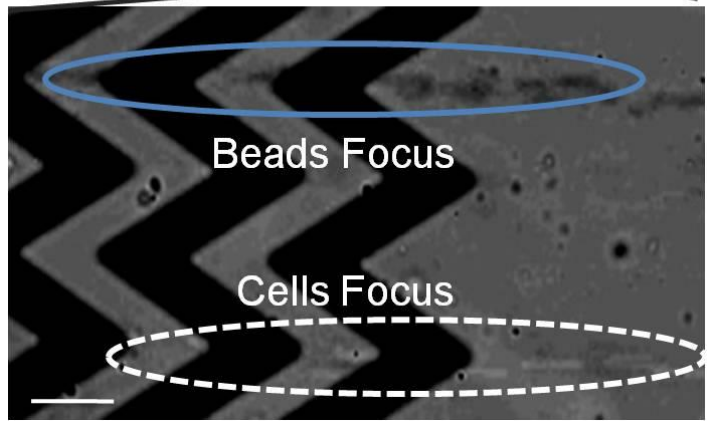

**(c)**

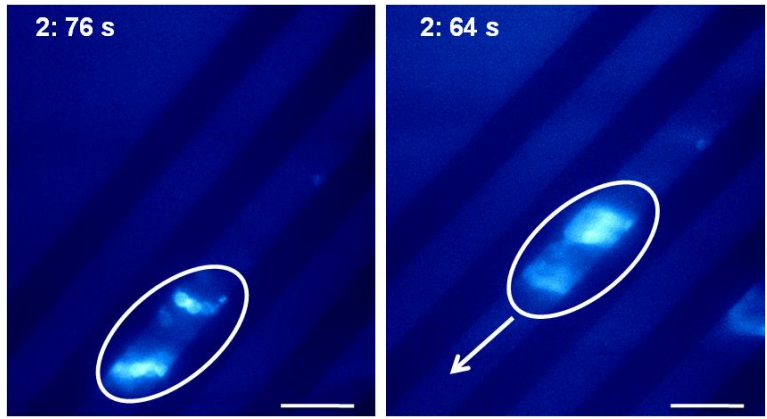

**(d)**

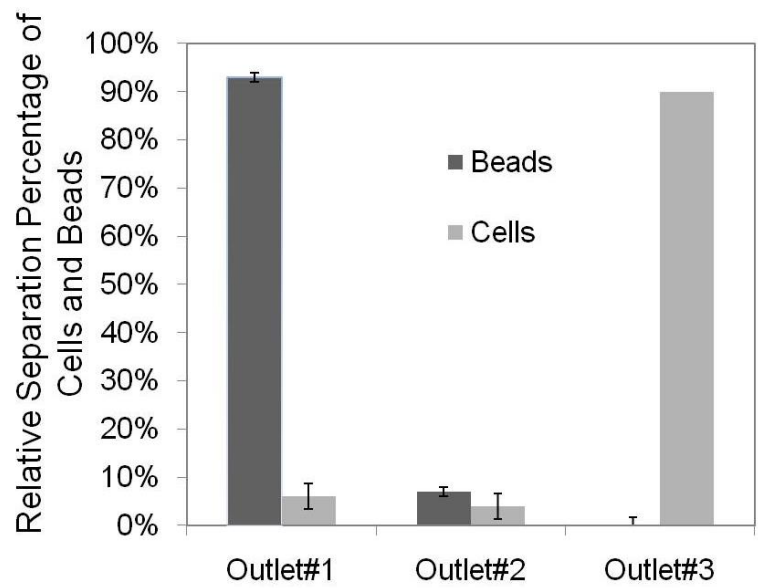

**Figure 5**

**(a)**

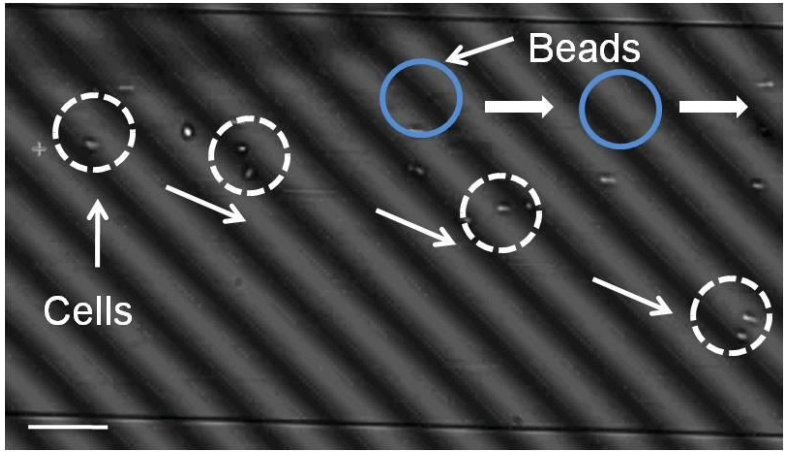

**(b)**

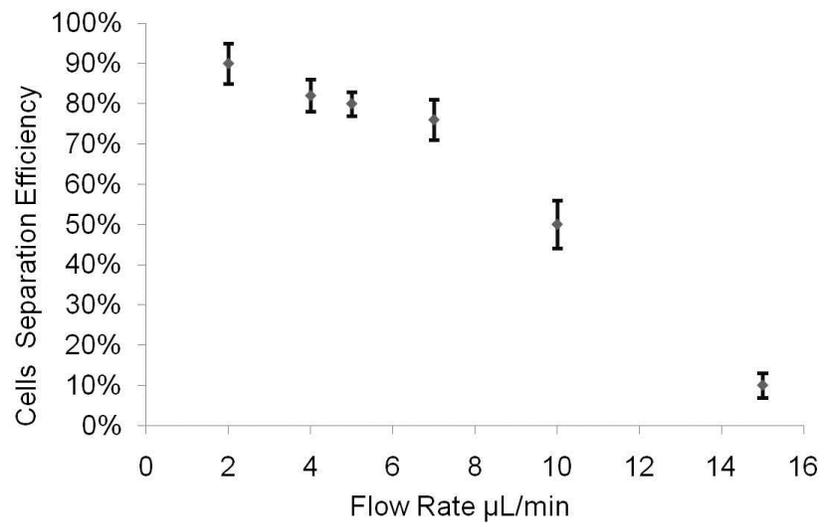

**(c)**

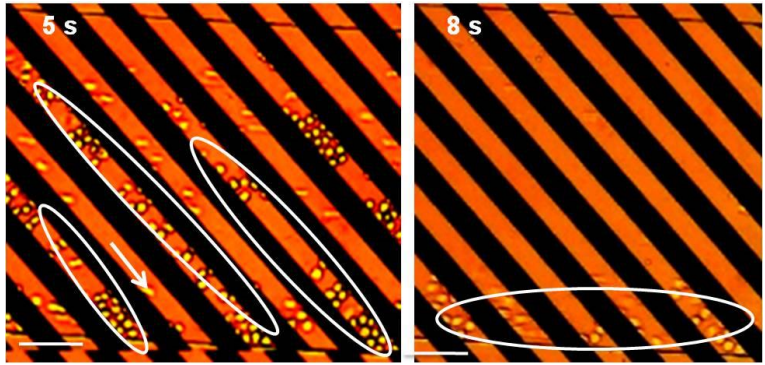

**(d)**

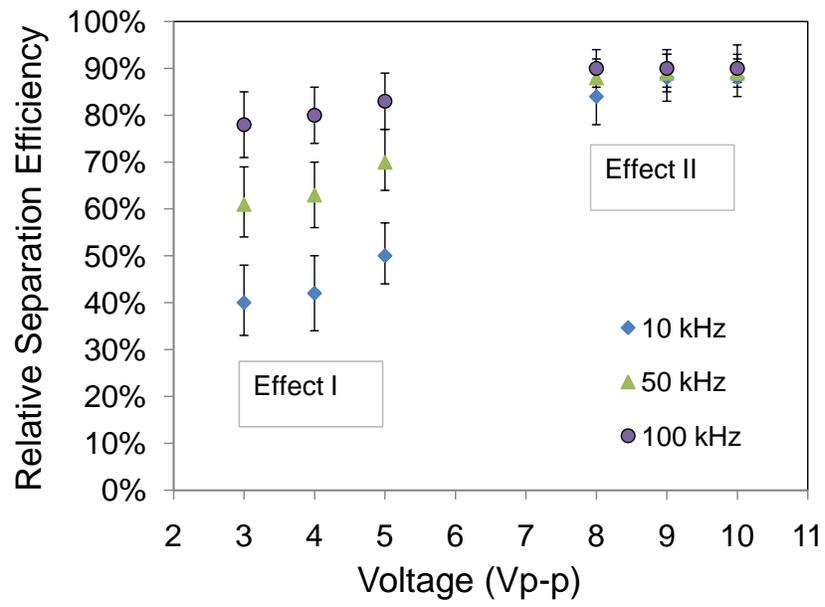